\documentclass[prd,twocolumn,amsmath,amssymb]{revtex4}
\usepackage{graphicx}%
\def\pp{\pi^+\pi^-}

\begin{document}

\title{
\boldmath Measurements of the observed cross sections for
$e^+e^-\rightarrow$ exclusive light hadrons containing $\pi^0\pi^0$
at $\sqrt s= 3.773$, 3.650 and 3.6648 GeV}

\author{
\small M.~Ablikim$^{1}$,              J.~Z.~Bai$^{1}$, Y.~Bai$^{1}$,
Y.~Ban$^{11}$, X.~Cai$^{1}$, H.~F.~Chen$^{15}$, H.~S.~Chen$^{1}$,
H.~X.~Chen$^{1}$, J.~C.~Chen$^{1}$, Jin~Chen$^{1}$
% ...
\\
\vspace{0.2cm}
(BES Collaboration)\\
\vspace{0.2cm}
{\it
$^{1}$ Institute of High Energy Physics, Beijing 100049, People's Republic of China\\
% ...
$^{11}$ Peking University, Beijing 100871, People's Republic of China\\
% ...
$^{15}$ University of Science and Technology of China, Hefei 230026,
People's Republic of China\\
% ...
}
}

\begin{abstract}
By analyzing the data sets of 17.3, 6.5 and 1.0 pb$^{-1}$ taken,
respectively, at $\sqrt s= 3.773$, 3.650 and 3.6648 GeV with the
BES-II detector at the BEPC collider, we measure the observed cross
sections for $e^+e^-\to \pi^+\pi^-\pi^0\pi^0$, $K^+K^-\pi^0\pi^0$,
$2(\pi^+\pi^-\pi^0)$, $K^+K^-\pi^+\pi^-\pi^0\pi^0$ and
$3(\pi^+\pi^-)\pi^0\pi^0$ at the three energy points. Based on these
cross sections we set the upper limits on the observed cross
sections and the branching fractions for $\psi(3770)$ decay into
these final states at 90\% C.L..
\end{abstract}

\maketitle
 \oddsidemargin
-0.2cm \evensidemargin -0.2cm

\section{Introduction}
In the past thirty years, it is expected that almost all of the
$\psi(3770)$ decay into $D\bar D$ meson pairs. However, the earlier
published data \cite{pdg04} show that the $\psi(3770)$ production
cross section exceeds the $D\bar D$ production cross section by
about 38\% \cite{hepex_0506051}. Recently, CLEO Collaboration
measured the cross section for $e^+e^-\to\psi(3770)\to$ non-$D\bar
D$ to be $(-0.01\pm0.08^{+0.41}_{-0.30})$ nb \cite{prl96_092002}.
However, by analyzing different data samples with different methods,
BES Collaboration extracted the branching fraction for
$\psi(3770)\to$ non$-D\bar D$ decay to be $(15\pm5)\%$
\cite{plb641_145,prl97_121801,plb659_74,prd76_000000,pdg07}, which
means that there may exist substantial non-$D\bar D$ final states in
the $\psi(3770)$ decays, or there are some new structure or physics
effects which may partially be responsible for the large ${\rm
non-}D\bar{D}$ branching fraction of the $\psi(3770)$ decays
measured by the BES collaboration \cite{prl101_102004,plb668_263}.

BES Collaboration found the first ${\rm non-}D\bar{D}$ decay mode of
$\psi(3770)$, i.e. $\psi(3770)\rightarrow J/\psi\pi^+\pi^-$ for the
first time, and extracted the branching fraction for $\psi(3770)\to
J/\psi\pi^+\pi^-$ to be $(0.34\pm0.14\pm0.09)\%$
\cite{hepnp28_325,plb605_63}. Latter, CLEO Collaboration confirmed
the BES observation \cite{prl96_082004} and observed more
$\psi(3770)$ exclusive non-$D\bar D$ decays, $\psi(3770)\to
J/\psi\pi^0\pi^0$, $J/\psi\pi^0$, $J/\psi\eta$ \cite{prl96_082004},
$\gamma\chi_{cJ}(J=0,1,2)$ \cite{prl96_182002,prd74_031106} and
$\phi\eta$ \cite{prd74_012005}, etc. However, the sum of these
measured branching fractions for $\psi(3770)\to$ exclusive
non-$D\bar D$ decays is not more than 2\%. BES and CLEO
Collaborations have also made many efforts to search for
$\psi(3770)$ exclusive charmless decays by comparing the cross
sections for $e^+e^-\to$ exclusive light hadrons measured at the
peak of $\psi(3770)$ and at some continuum energy points
\cite{prd70_077101,prd72_072007,plb650_111,plb656_30,epjc52_805}
\cite{prd74_012005,prl96_032003,prd73_012002}.

In this Letter, we report another effort to search for $\psi(3770)$
exclusive charmless decays, which are performed by studying some
processes containing $\pi^0\pi^0$ mesons in the final states with
the same method as the one used in Refs.
\cite{plb650_111,plb656_30,epjc52_805}. We measure the observed
cross sections for $e^+e^-\to \pi^+\pi^-\pi^0\pi^0$,
$K^+K^-\pi^0\pi^0$, $2(\pi^+\pi^-\pi^0)$,
$K^+K^-\pi^+\pi^-\pi^0\pi^0$ and $3(\pi^+\pi^-)\pi^0\pi^0$ at $\sqrt
s = 3.773$, 3.650 and 3.6648 GeV. Then, we set the upper limits on
the observed cross sections and the branching fractions for
$\psi(3770)$
decay into these final states at 90\% C.L..
The data used in the analysis were taken at the center-of-mass
energies $\sqrt{s} = 3.773$, 3.650 and 3.6648 GeV with the BES-II
detector at the BEPC collider, corresponding to the integrated
luminosities of 17.3, 6.5 and 1.0 pb$^{-1}$, respectively. In the
Letter, we call, respectively, the three data sets as the
$\psi(3770)$ resonance data, the continuum data 1 and the continuum
data 2.

\section{BES-II detector}
The BES-II is a conventional cylindrical magnetic detector \cite
{nima344_319,nima458_627} operated at the Beijing Electron-Positron
Collider (BEPC). A 12-layer vertex chamber (VC) surrounding the
beryllium beam pipe provides input to the event trigger, as well as
coordinate information. A forty-layer main drift chamber (MDC)
located just outside the VC yields precise measurements of charged
particle trajectories with a solid angle coverage of $85\%$ of
4$\pi$; it also provides ionization energy loss ($dE/dx$)
measurements which are used for particle identification. Momentum
resolution of $1.7\%\sqrt{1+p^2}$ ($p$ in GeV/$c$) and $dE/dx$
resolution of $8.5\%$ for Bhabha scattering electrons are obtained
for the data taken at $\sqrt s= 3.773$ GeV. An array of 48
scintillation counters surrounding the MDC measures the time of
flight (TOF) of charged particles with a resolution of about 180 ps
for electrons. Outside the TOF, a 12 radiation length, lead-gas
barrel shower counter (BSC), operating in limited streamer mode,
measures the energies of electrons and photons over $80\%$ of the
total solid angle with an energy resolution of
$\sigma_E/E=0.22/\sqrt{E}$ ($E$ in GeV) and spatial resolutions of
$\sigma_{\phi}=7.9$ mrad and $\sigma_z=2.3$ cm for electrons. A
solenoidal magnet outside the BSC provides a 0.4 T magnetic field in
the central tracking region of the detector. Three double-layer muon
counters instrument the magnet flux return and serve to identify
muons with momentum greater than 0.5 GeV/c. They cover $68\%$ of the
total solid angle.

\section{Event selection}
\label{evtsel}

The exclusive light hadron final states of $\pi^+\pi^-\pi^0\pi^0$,
$K^+K^-\pi^0\pi^0$, $2(\pi^+\pi^-\pi^0)$,
$K^+K^-\pi^+\pi^-\pi^0\pi^0$ and $3(\pi^+\pi^-)\pi^0\pi^0$ are
studied by examining the different photon combinations from
$\pi^+\pi^-\gamma_1\gamma_2\gamma_3\gamma_4$,
$K^+K^-\gamma_1\gamma_2\gamma_3\gamma_4$,
$2(\pi^+\pi^-)\gamma_1\gamma_2\gamma_3\gamma_4$,
$K^+K^-\pi^+\pi^-\gamma_1\gamma_2\gamma_3\gamma_4$ and
$3(\pi^+\pi^-)\gamma_1\gamma_2\gamma_3\gamma_4$, respectively. The
$\pi^0$ mesons are reconstructed through the decay
$\pi^0\to\gamma\gamma$.

For each candidate event, it is required that there are at least two
charged tracks to be well reconstructed in the MDC with good helix
fits and at least four neutral tracks to be well reconstructed in
the BSC. The charged track is required to have the polar angle
satisfying $|\cos\theta|<0.85$, and originate from the interaction
region $V_{xy}<2.0$ cm and $|V_{z}|<20.0$ cm. Here, $V_{xy}$ and
$|V_{z}|$ are the closest approaches of the charged track in the
$xy$ plane and in the $z$ direction.

The charged particles are identified by using the $dE/dx$ and TOF
measurements, with which the combined confidence levels $CL_{\pi}$
and $CL_{K}$ for pion and kaon hypotheses are calculated. The
candidate charged tracks satisfying $CL_{\pi}>$0.001 and
$CL_{K}>CL_{\pi}$ are identified as pion and kaon, respectively.

The good photons are selected with the BSC measurements. They are
required to satisfy the following criteria: the energy deposited in
the BSC is greater than 50 MeV, the electromagnetic shower starts in
the first 5 layers, the angle between the photon and the nearest
charged track is greater than 22$^\circ$ \cite{plb597_39} and the
opening angle between the cluster development direction and the
photon emission direction is less than 37$^\circ$ \cite{plb597_39}.

For each selected candidate event, there may be several different
charged or neutral track combinations satisfying the above selection
criteria for exclusive light hadron final states. Each combination
is subjected to an energy-momentum conservative kinematic fit.
Candidates with a fit probability larger than 1$\%$ are accepted. If
more than one combination satisfies the selection criteria in an
event, only the combination with the largest fit probability is
retained.

At center-of-mass energy of 3.773 GeV, the events of
$\psi(2S)\rightarrow(\gamma)J/\psi\pi^0\pi^0$, with $J/\psi\to
e^+e^-$ or $J/\psi\to\mu^+\mu^-$ may be misidentified as the
$\pi^+\pi^-\pi^0\pi^0$ final state due to misidentifying $e^+e^-$ or
$\mu^+\mu^-$ pair as $\pi^+\pi^-$ pair. We suppress these events by
requiring the invariant mass of $\pi^+\pi^-$ combination from each
selected $\pi^+\pi^-\pi^0\pi^0$ candidate event to be less than 3.0
GeV/c$^2$.

The $K^+K^-\pi^+\pi^-\pi^0\pi^0$ final state suffers more
contaminations from $D\bar D$ decays than the other four modes. To
eliminate these contaminations, we exclude the events from $D\bar D$
decays by rejecting those in which the $D$ and $\bar D$ mesons can
be reconstructed in the decay modes of $D^0\to K^-\pi^+\pi^0$ and
$\bar{D^0}\to K^+\pi^-\pi^0$ \cite{npb727_395}. The residual
contaminations from the other $D$ meson decays are further removed
based on Monte Carlo simulation.

\section{Data Analysis}
\label{sec:data_analysis}

\begin{figure}[htbp]
\begin{center}
  \includegraphics[width=8cm]
{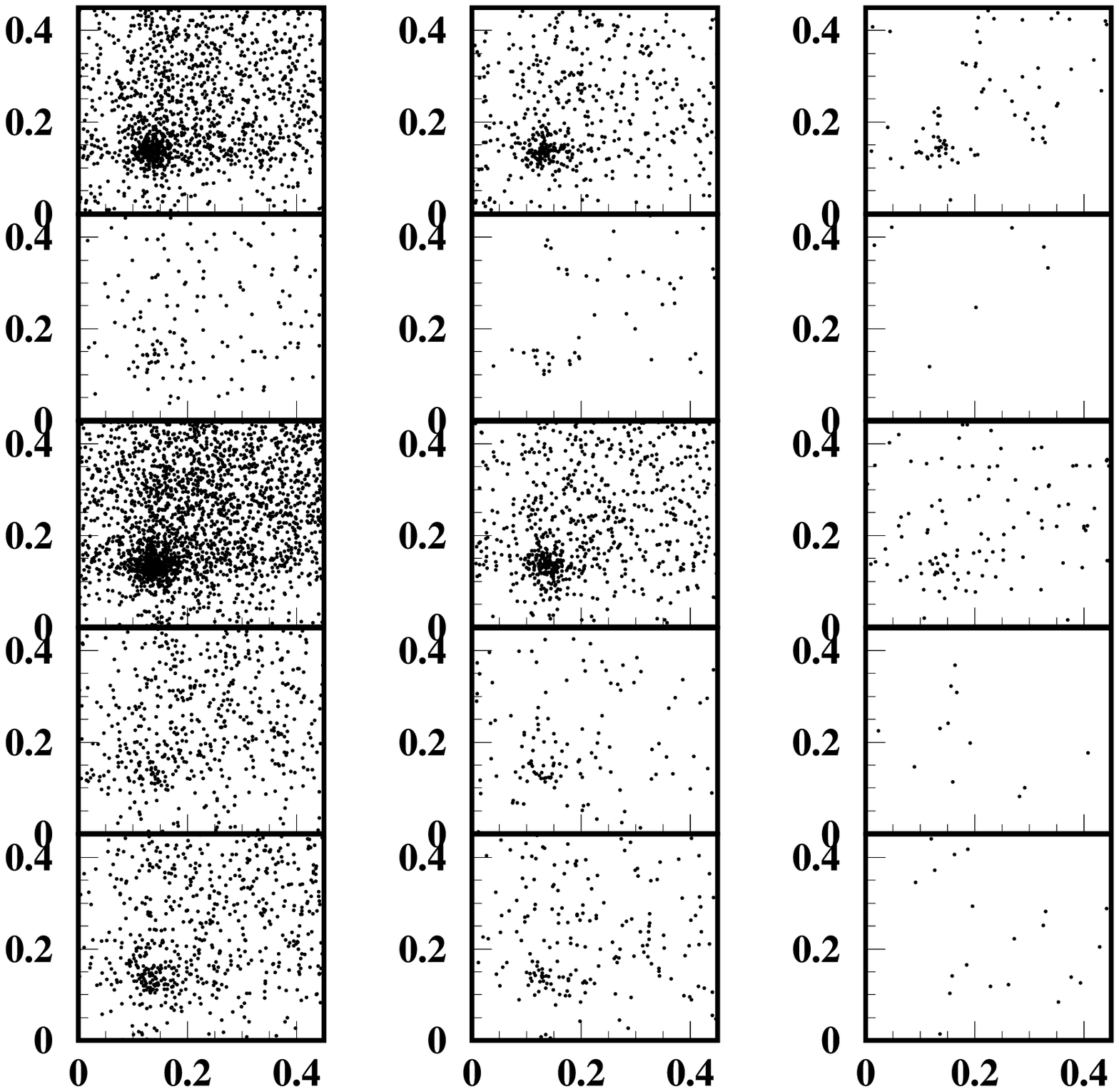}
\put(-145,-15) {\normalsize \bf $M_{\gamma_{i}\gamma_{j}}$(GeV/$c^2$)}
\put(-245,70) {\normalsize \rotatebox{90}
{\bf $M_{\gamma_{i^{\prime}}\gamma_{j^{\prime}}}$(GeV/$c^2$)}}
\put(-180,205) {\normalsize \bf (a)}
\put(-100,205) {\normalsize \bf (${\rm a^{\prime}}$)}
\put(-23,205) {\normalsize \bf (${\rm a^{\prime\prime}}$)}
\put(-180,162) {\normalsize \bf (b)}
\put(-100,162) {\normalsize \bf (${\rm b^{\prime}}$)}
\put(-23,162) {\normalsize \bf (${\rm b^{\prime\prime}}$)}
\put(-180,120) {\normalsize \bf (c)}
\put(-100,120) {\normalsize \bf (${\rm c^{\prime}}$)}
\put(-23,120) {\normalsize \bf (${\rm c^{\prime\prime}}$)}
\put(-180,77) {\normalsize \bf (d)}
\put(-100,77) {\normalsize \bf (${\rm d^{\prime}}$)}
\put(-23,77) {\normalsize \bf (${\rm d^{\prime\prime}}$)}
\put(-180,37) {\normalsize \bf (e)}
\put(-100,37) {\normalsize \bf (${\rm e^{\prime}}$)}
\put(-23,37) {\normalsize \bf (${\rm e^{\prime\prime}}$)}
\caption {The scatter plots of
$M_{\gamma_{i^{\prime}}\gamma_{j^{\prime}}}$ versus $M_{\gamma_i\gamma_j}$
of the candidates for $e^+e^-\to$
(a) $\pi^+\pi^-\pi^0\pi^0$,
(b) $K^+K^-\pi^0\pi^0$,
(c) $2(\pi^+\pi^-\pi^0)$,
(d) $K^+K^-\pi^+\pi^-\pi^0\pi^0$ and
(e) $3(\pi^+\pi^-)\pi^0\pi^0$
selected from the $\psi(3770)$ resonance data (left),
the continuum data 1 (middle) and
the continuum data 2 (right).}
\label{fig:dot}
\end{center}
\end{figure}

For each $\gamma_1\gamma_2\gamma_3\gamma_4$ combination satisfying
the above selection criteria, there are three possible photon
combinations ([$\pi^0_{\gamma_1\gamma_2}\pi^0_{\gamma_3\gamma_4}$],
[$\pi^0_{\gamma_1\gamma_3}\pi^0_{\gamma_2\gamma_4}$],[$\pi^0_{\gamma_1\gamma_4}\pi^0_{\gamma_2\gamma_3}$])
which may be reconstructed as $\pi^0\pi^0$ meson pairs. We calculate
the invariant masses for the $\gamma_i\gamma_j$ and
$\gamma_{i^\prime}\gamma_{j^{\prime}}$ combinations,
$M_{\gamma_i\gamma_j}$ and
$M_{\gamma_{i^{\prime}}\gamma_{j^{\prime}}}$($i$, $j$, $i^{\prime}$,
$j^{\prime}$ = 1, or 2, or 3, or 4; $i < j$, $i^{\prime} <
j^{\prime}$, $i < i^{\prime}$ and $j < j^{\prime}$) with the fitted
momentum vectors from the kinematic fit.
Figure \ref{fig:dot} shows the scatter plot for
$M_{\gamma_{i^{\prime}}\gamma_{j^{\prime}}}$ versus
$M_{\gamma_i\gamma_j}$ of the candidates for the
$\pi^+\pi^-\pi^0\pi^0$, $K^+K^-\pi^0\pi^0$, $2(\pi^+\pi^-\pi^0)$,
$K^+K^-\pi^+\pi^-\pi^0\pi^0$ and $3(\pi^+\pi^-)\pi^0\pi^0$ final
states selected from the data including the three possible
combinations.

\begin{figure}[htbp]
\begin{center}
  \includegraphics[width=8cm]
{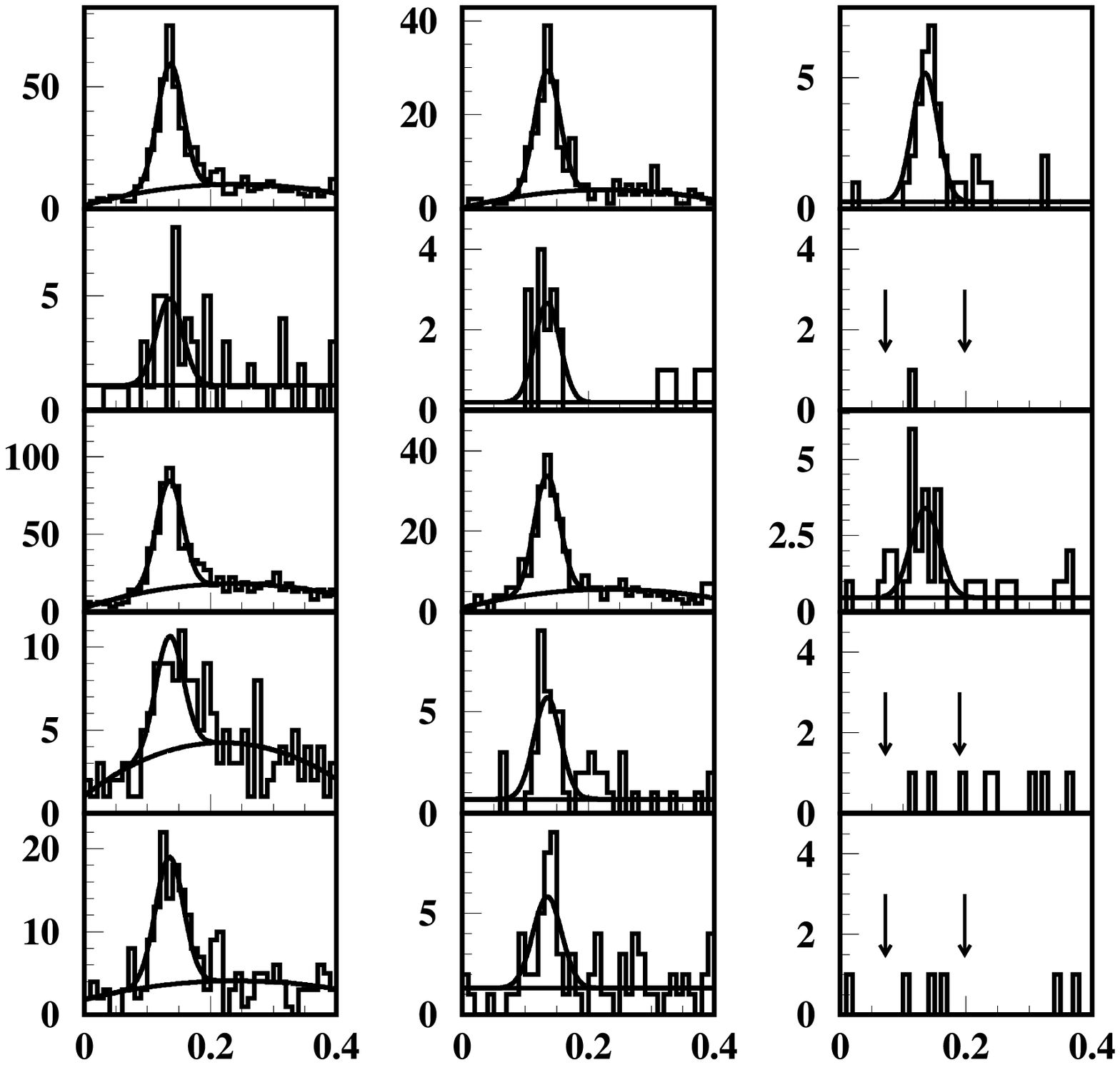} \put(-145,-15){\normalsize \bf
$M_{\gamma_{i}\gamma_{j}}$(GeV/$c^2$)} \put(-245,50) {\normalsize
\rotatebox{90} {\bf Events(/0.01 GeV/$c^2$)}} \put(-176,200)
{\normalsize \bf (a)} \put(-102,200) {\normalsize \bf (${\rm
a^{\prime}}$)} \put(-28,200) {\normalsize \bf (${\rm
a^{\prime\prime}}$)} \put(-176,159) {\normalsize \bf (b)}
\put(-102,159) {\normalsize \bf (${\rm b^{\prime}}$)} \put(-28,159)
{\normalsize \bf (${\rm b^{\prime\prime}}$)} \put(-176,119)
{\normalsize \bf (c)} \put(-102,119) {\normalsize \bf (${\rm
c^{\prime}}$)} \put(-28,119) {\normalsize \bf (${\rm
c^{\prime\prime}}$)} \put(-176,77) {\normalsize \bf (d)}
\put(-102,77) {\normalsize \bf (${\rm d^{\prime}}$)} \put(-28,77)
{\normalsize \bf (${\rm d^{\prime\prime}}$)} \put(-176,38)
{\normalsize \bf (e)} \put(-102,38) {\normalsize \bf (${\rm
e^{\prime}}$)} \put(-28,38) {\normalsize \bf (${\rm
e^{\prime\prime}}$)} \caption { The distributions of the invariant
masses for the $\gamma_i\gamma_j$ combinations from the candidates
for $e^+e^-\to$ (a) $\pi^+\pi^-\pi^0\pi^0$, (b) $K^+K^-\pi^0\pi^0$,
(c) $2(\pi^+\pi^-\pi^0)$, (d) $K^+K^-\pi^+\pi^-\pi^0\pi^0$ and (e)
$3(\pi^+\pi^-)\pi^0\pi^0$ from the $\psi(3770)$ resonance data
(left), the continuum data 1 (middle) and the continuum data 2
(right), where the pairs of arrows show the $\pi^0$ signal region.}
\label{fig:inv}
\end{center}
\end{figure}

In Fig. \ref{fig:dot}, the cluster around the $\pi^0\pi^0$ signal
region indicates the production of the process containing
$\pi^0\pi^0$ mesons in the final state. In the following analysis,
the mass window of $\pm3\sigma_{M_{\gamma\gamma}}$($\pm$60
MeV/$c^2$) around the $\pi^0$ nominal mass is taken as the $\pi^0$
signal region, where $\sigma_{M_{\gamma\gamma}}$ is the $\pi^0$ mass
resolution determined by Monte Carlo simulation. In each figure,
projecting the events with
$M_{\gamma_{i^{\prime}}\gamma_{j^{\prime}}}$ in the $\pi^0$ signal
region onto $M_{\gamma_{i}\gamma_{j}}$ axis, we obtain the
distribution of the $\gamma_i\gamma_j$ invariant masses for each
process as shown in Fig. \ref{fig:inv}. Fitting each of the
$\gamma_i\gamma_j$ invariant mass spectra in Fig. \ref{fig:inv} with
a Gaussian function for the $\pi^0$ signal and a polynomial for the
background yields the number $N_{\rm obs}^{\pi^0_{\gamma_i\gamma_j}}$
of the events for each process observed from each of the data sets.
However, in each event, there may be more than one combination entering
the $\gamma_i\gamma_j$ invariant mass spectra. The number $N^{\rm rc}$
of the repeated counting events in the observed number
$N_{\rm obs}^{\pi^0_{\gamma_i\gamma_j}}$ of $\pi^0$s from the fit is
subtracted from the number of
$N_{\rm obs}^{\pi^0_{\gamma_i\gamma_j}}$. The number of $N^{\rm rc}$
is accounted via the number of $N_{\rm obs}^{\pi^0_{\gamma_i\gamma_j}}$,
the number of the repeated counting events and the number of all the
events in the $\pi^0$ signal region, as well as the repeated counting
rate of the combinatorial background events which is estimated with
the events in sideband region.

Using the number of events observed in a sideband region which is
the mass window of $\pm60$ MeV/$c^2$ around 0.335 GeV$/c^2$ in the
$\gamma_i^{\prime}\gamma_j^{\prime}$ invariant mass spectra, we can
estimate the contributions of the combinatorial
$\gamma_{i^\prime}\gamma_{j^\prime}$ background in the
$\pi^0_{\gamma_{i^\prime}\gamma_{j^\prime}}$ signal region.
We obtain the number of these contributions,
$N^{\pi^0_{\gamma_i\gamma_j}}_{\rm sid}$, by
normalizing the fitted number of $\pi^0$s observed in the sideband region.
Here, the normalization factor is the ratio of the number of the background
events in the signal region over the number of the background events in the
sideband region.

However, there are only a few events in Fig. \ref{fig:inv}(${\rm
b^{\prime\prime}}$), (${\rm d^{\prime\prime}}$) and (${\rm
e^{\prime\prime}}$). Counting the events with $M_{\gamma\gamma}$
within the $\pi^0$ signal region, we obtain 1, 2 and 3 candidate
events for $e^+e^-\to K^+K^-\pi^0\pi^0$,
$K^+K^-\pi^+\pi^-\pi^0\pi^0$ and $3(\pi^+\pi^-)\pi^0\pi^0$ from the
continuum data 2, respectively. In this case, we set the upper limit
$N^{\rm up}$ on the number of the signal events at 90\% confidence
level (C.L.). Here, we use the Feldman-Cousins method
\cite{prd57_3873} and ignore the background.

\section{Other background subtraction}

In Section \ref{sec:data_analysis}, we have considered the
combinatorial $\gamma\gamma$ background in the $\pi^0\pi^0$
reconstruction. However, there are still some other kinds of
backgrounds from $J/\psi$ and $\psi(3686)$ decays due to ISR
returns, from the other final states due to the misidentification
between charged pion and kaon, from the decays of $\psi(3770) \to
J/\psi\pp$, $J/\psi\pi^0\pi^0$, $J/\psi\pi^0$, $\gamma \chi_{cJ}
\hspace{0.1cm}(J=0,1,2)$, and from $D\bar D$ decays. In order to
directly compare the cross sections for $e^+e^-\to f$ ($f$
represents exclusive light hadron final state) measured at the peak
of $\psi(3770)$ and at some continuum energy points, we need to
remove these background events. The method of background subtraction
has been described in detail in Ref. \cite{plb650_111}. Monte Carlo
study shows that the contaminations from the decays of $\psi(3770)
\to J/\psi\pp$, $J/\psi\pi^0\pi^0$, $J/\psi\pi^0$ and $\psi(3770)
\to\gamma \chi_{cJ}$ can be neglected \cite{plb650_111} and the
contaminations from the final states with an extra photon can be
ignored.

For the process of $K^+K^-\pi^+\pi^-\pi^0\pi^0$, even though we have
removed the main contaminations from $D\bar D$ decays in the
previous event selection (see section \ref{evtsel}), there are still
some residual contaminations from $D\bar D$ decays, which can
satisfy the selection criteria for the $K^+K^-\pi^+\pi^-\pi^0\pi^0$
final state. The number of these residual contaminations are further
removed based on Monte Carlo simulation.

Since we don't know the details about the resonance and the
continuum amplitudes, we neglect the possible interference between
them in the following analysis. In this case, by subtracting $N^{\rm
rc}$, $N^{\pi^0_{\gamma_i\gamma_j}}_{\rm sid}$ and the number
$N^{\rm b}$ of these contaminations from the number
$N^{\pi^0_{\gamma_i\gamma_j}}_{\rm obs}$ of the candidate events, we
obtain the net numbers $N^{\rm net}$ of the signal events for each
process. The numbers of $N^{\pi^0_{\gamma_i\gamma_j}}_{\rm obs}$,
$N^{\rm rc}$, $N^{\pi^0_{\gamma_i\gamma_j}}_{\rm sid}$, $N^{\rm b}$
and $N^{\rm net}$ (or $N^{\rm up}$) are summarized in Tabs.
\ref{tab:crs3773}, \ref{tab:crs3650} and \ref{tab:crs36648}.

\section{Results}

\subsection{Monte Carlo efficiency}

We estimate the detection efficiency $\epsilon$ for $e^+e^-\to f$ by
using a phase space generator including initial state radiation and
vacuum polarization corrections \cite{yf41_377} with $1/s$ cross
section energy dependence. Final state radiation \cite{cpc79_291}
decreases the detection efficiency not more than 0.5\%. The Monte
Carlo events are generated based on the Monte Carlo simulation for
the BES-II detector \cite{nima552_344}. By analyzing these Monte
Carlo events, we obtain the detection efficiency for each process at
each center-of-mass energy. They are summarized in the fifth columns
of Tabs. \ref{tab:crs3773}, \ref{tab:crs3650} and
\ref{tab:crs36648}. The detection efficiencies listed in the tables
do not include the branching fraction for $\pi^0\to\gamma\gamma$.

\subsection{Observed cross sections}

The observed cross section for $e^+e^- \to f$ is obtained by
dividing the number $N^{\rm net}$ of the signal events by the
integrated luminosity $\mathcal{L}$ of the data set, the detection
efficiency $\epsilon$ and the square of the branching fraction
${\mathcal B}^2(\pi^0\to\gamma\gamma)$ for $\pi^0\to\gamma\gamma$,
\begin{equation}
\sigma_{e^+e^-\to f} = \frac{N^{\rm net}} {\mathcal{L}
\times \epsilon
\times {\mathcal B}^2(\pi^0\to\gamma\gamma)
}.
\label{eq:crs}
\end{equation}
Inserting the numbers of $N^{\rm net}$, $\mathcal{L}$, $\epsilon$
and ${\mathcal B}(\pi^0\to\gamma\gamma)$ in Eq. (\ref{eq:crs}), we
obtain $\sigma_{e^+e^-\to f}$ for each process at $\sqrt s= 3.773$,
3.650 and 3.6648 GeV, respectively. They are summarized in the last
columns of Tabs. \ref{tab:crs3773}, \ref{tab:crs3650} and
\ref{tab:crs36648}, where the first error is statistical and the
second systematic. The systematic error in the cross section
measurement arises mainly from the uncertainties in integrated
luminosity of the data set ($\sim2.1\%$
\cite{plb641_145,prl97_121801}), photon selection ($\sim2.0\%$, per
photon), tracking efficiency ($\sim2.0\%$ per track), particle
identification ($\sim0.5\%$ per pion or kaon), kinematic fit
($\sim1.5\%$), Monte Carlo statistics ($\sim(3.3\sim5.7)\%$),
branching fraction quoted from PDG ($\sim0.03\%$ for $B(\pi^0\to
\gamma\gamma)$), background subtraction ($\sim(0.0\sim6.1)\%$),
fitting to the mass spectra ($\sim(2.4\sim9.5)\%$), and Monte Carlo
modeling ($\sim6.0\%$). Adding these uncertainties in quadrature
yields the total systematic error $\Delta_{\rm sys}$ for each
process at each center-of-mass energy.

\subsection{Upper limits on the observed cross sections}

The upper limits on the observed cross sections for $e^+e^-\to
K^+K^-\pi^0\pi^0$, $K^+K^-\pi^+\pi^-\pi^0\pi^0$ and
$3(\pi^+\pi^-)\pi^0\pi^0$ at $\sqrt s=3.6648$ GeV are set by
\begin{eqnarray}
\sigma^{\rm up}_{e^+e^-\to f}=
\frac{N^{\rm up}} {\mathcal{L} \times
\epsilon
\times (1-\Delta_{\rm sys})
\times {\mathcal B}^2(\pi^0\to\gamma\gamma)
},
\label{eq:crsup}
\end{eqnarray}
where $N^{\rm up}$ is the upper limit on the number of the signal
events setting based on Feldman-Cousins method \cite{prd57_3873},
and $\Delta_{\rm sys}$ is the systematic error in the cross section
measurement. Inserting the numbers of $N^{\rm up}$, $\mathcal{L}$,
$\epsilon$, $\Delta_{\rm sys}$ and ${\mathcal
B}(\pi^0\to\gamma\gamma)$ in Eq. (\ref{eq:crsup}), we obtain
$\sigma^{\rm up}_{e^+e^-\to f}$ for these processes at $\sqrt s=
3.6648$ GeV, which are also summarized in the last column of Tab.
\ref{tab:crs36648}.

\subsection{Upper limits on the observed cross sections and the
branching fractions for $\psi(3770)\to f$}

In the $\psi(3770)$ resonance region \footnote { Assuming that there
is only one $\psi(3770)$ in the energy region from 3.70 to 3.87
GeV.}, if we ignore the possible interference effects between the
continuum and resonance amplitudes and the difference of the vacuum
polarization corrections at $\sqrt s= 3.773$ and 3.650 GeV, the
contributions of the continuum production for $e^+e^- \to f$ at
$\sqrt s= 3.773$ GeV can be expected by these measured at $\sqrt s=
3.650$ GeV. In this case, the observed cross section for
$\psi(3770)\to f$ at $\sqrt s= 3.773$ GeV can be written as
\begin{equation}
\sigma_{\psi(3770)\to f}=
\sigma^{\rm 3.773\hspace{0.05cm}GeV}_{e^+e^-\to f} -
f_{\rm co}\times \sigma^{\rm 3.650\hspace{0.05cm}GeV}_{e^+e^- \to f},
\label{eq:obscrs}
\end{equation}
where $\sigma^{\rm 3.773\hspace{0.05cm}GeV}_{e^+e^- \to f}$ and
$\sigma^{\rm 3.650 \hspace{0.05cm}GeV}_{e^+e^- \to f}$ are the
observed cross sections for $e^+e^- \to f$ measured at $\sqrt s=
3.773$ and 3.650 GeV, respectively; $f_{\rm co}$ is the factor
accounting for the 1/s dependence of the cross section. Neglecting
the difference in the corrections for ISR and vacuum polarization
effects at the two energy points, we have $f_{\rm co} =
(3.650/3.773)^2$. With the $\sigma^{\rm
3.773\hspace{0.05cm}GeV}_{e^+e^- \to f}$ and $\sigma^{\rm 3.650
\hspace{0.05cm}GeV}_{e^+e^- \to f}$ listed in Tabs.
\ref{tab:crs3773} and \ref{tab:crs3650}, we determine
$\sigma_{\psi(3770)\to f}$ at $\sqrt s= 3.773$ GeV for each process.
They are summarized in the second column of Tab. \ref{tab:up_psipp},
where the first error is the statistical, the second is the
independent systematic arising from the uncertainties in the Monte
Carlo statistics, in the fit to the mass spectrum and in the
background subtraction, and the third is the common systematic error
arising from the other uncertainties as discussed in the subsection
B.

Assuming that the cross section for $\psi(3770) \to f$ follows a
Gaussian distribution, we set the upper limit on the observed cross
section for $\psi(3770)\to f$ at $\sqrt s= 3.773$ GeV, $\sigma^{\rm
up}_{\psi(3770) \to f}$, which are summarized in the third column of
Tab. \ref{tab:up_psipp}.

The upper limit on the branching fraction ${\mathcal B}^{\rm
up}_{\psi(3770)\to f}$ for $\psi(3770)\to f$ is set by dividing its
upper limit on the observed cross section $\sigma^{\rm
up}_{\psi(3770) \to f}$ by the observed cross section $\sigma^{\rm
obs}_{\psi(3770)}=(7.15\pm0.27\pm0.27)$ nb
\cite{plb650_111,prl97_121801,plb652_238} for the $\psi(3770)$
production at $\sqrt s= 3.773$ GeV and a factor $(1-\Delta
\sigma^{\rm obs}_{\psi(3770)})$, where $\Delta \sigma^{\rm
obs}_{\psi(3770)}$ is the relative error of the
$\sigma^{\rm obs}_{\psi(3770)}$.
The results on ${\mathcal B}^{\rm up}_{\psi(3770)\to f}$ are
summarized in the last column of Tab. \ref{tab:up_psipp}.

\section {\bf Summary}

In summary, by analyzing the data sets of 17.3, 6.5 and 1.0
pb$^{-1}$ taken, respectively, at $\sqrt s= 3.773$, 3.650 and 3.6648
GeV with the BES-II detector at the BEPC collider, we measured the
observed cross sections for $e^+e^-\to \pi^+\pi^-\pi^0\pi^0$,
$K^+K^-\pi^0\pi^0$, $2(\pi^+\pi^-\pi^0)$,
$K^+K^-\pi^+\pi^-\pi^0\pi^0$ and $3(\pi^+\pi^-)\pi^0\pi^0$ at the
three energy points. Based on the measured cross sections for these
processes at $\sqrt s = 3.773$ and 3.650 GeV, as well as the
observed cross section for the $\psi(3770)$ production at $\sqrt s=
3.773$ GeV, we also set the upper limits on the observed cross
sections and the branching fractions for $\psi(3770)$ decay into
these final states at $90\%$ C.L.. These measurements provide useful
experimental information for better understanding the possible
excess of the cross section for the $\psi(3770)$ production relative
to the cross section for the $D\bar D$ production and the mechanism
of the continuum light hadron production.

\section{Acknowledgments}
The BES collaboration thanks the staff of BEPC for their hard
efforts. This work is supported in part by the National Natural
Science Foundation of China under contracts Nos. 10491300, 10225524,
10225525, 10425523, the Chinese Academy of Sciences under contract
No. KJ 95T-03, the 100 Talents Program of CAS under Contract Nos.
U-11, U-24, U-25, the Knowledge Innovation Project of CAS under
Contract Nos. U-602, U-34 (IHEP), the National Natural Science
Foundation of China under Contract  No. 10225522 (Tsinghua
University).

\begin{table*}[htbp]
\caption{The observed cross sections for $e^+e^-\to$ exclusive light
hadrons at $\sqrt s= 3.773$ GeV, where
$N^{\pi^0_{\gamma_i\gamma_j}}_{\rm obs}$ is the fitted number of
events with $M_{\gamma_{i'}\gamma_{j'}}$ in the $\pi^0$ signal
region observed from the $\psi(3770)$ resonance data, $N^{\rm rc}$
is the normalized number of the repeated counting events,
$N^{\pi^0_{\gamma_i\gamma_j}}_{\rm sid}$ is the normalized number of
events with $M_{\gamma_{i'}\gamma_{j'}}$ in the $\pi^0$ sideband
region, $N^{\rm b}$ is the total number of the background events,
$N^{\rm net}$ is the number of the signal events, $\epsilon$ is the
detection efficiency, $\Delta_{\rm sys}$ is the relative systematic
error in the cross section measurement, $\sigma$ is the observed
cross section.}
\begin{center}
\begin{tabular}{l|c|c|c|c|c|c|c|c}
\hline
$e^+e^-\to f$ & $N^{\pi^0_{\gamma_i\gamma_j}}_{\rm obs}$ & $N^{\rm rc}$ &
$N^{\pi^0_{\gamma_i\gamma_j}}_{\rm sid}$ & $N^{\rm b}$ & $N^{\rm net}$ & $\epsilon(\%)$ & $\Delta_{\rm sys}(\%)$ & $\sigma^{\rm obs}$ [pb] \\
\hline
$\pi^+\pi^-\pi^0\pi^0$ & $259.7\pm21.8$ & $5.2\pm7.2$ & $29.8\pm10.4$ & $7.1\pm2.2$ & $217.6\pm25.3$ & $6.00\pm0.21$ & 13.0 & $214.8\pm25.0\pm27.9$\\
$K^+K^-\pi^0\pi^0$ & $19.8\pm5.6$ & $0.0\pm0.0$ & $4.0\pm3.1$ & $2.0\pm0.5$ & $13.8\pm6.4$ & $3.06\pm0.15$ & 15.1 & $26.7\pm12.4\pm4.0$\\
$2(\pi^+\pi^-\pi^0)$ & $374.6\pm29.0$ & $31.5\pm10.6$ & $29.2\pm15.6$ & $8.5\pm1.4$ & $305.4\pm34.6$ & $1.72\pm0.07$ & 14.1 & $1051.5\pm119.2\pm148.3$\\
$K^+K^-\pi^+\pi^-\pi^0\pi^0$ & $38.2\pm9.5$ & $7.1\pm4.3$ & $0.7\pm4.3$ & $5.9\pm1.5$ & $24.5\pm11.4$ & $0.78\pm0.03$ & 16.8 & $186.0\pm86.4\pm31.2$\\
$3(\pi^+\pi^-)\pi^0\pi^0$ & $81.2\pm14.2$ & $17.0\pm5.1$ & $2.4\pm5.2$ & $2.6\pm0.7$ & $59.2\pm16.0$ & $0.36\pm0.02$ & 18.5 & $973.8\pm262.8\pm180.2$ \\
\hline
\end{tabular}
\label{tab:crs3773}
\end{center}
\end{table*}

\begin{table*}[htbp]
\caption{The observed cross sections for $e^+e^-\to$ exclusive light
hadrons at $\sqrt s= 3.650$ GeV, where
$N^{\pi^0_{\gamma_i\gamma_j}}_{\rm obs}$ is the fitted number of events
with $M_{\gamma_{i'}\gamma_{j'}}$ in the $\pi^0$ signal region
observed from the continuum data 1, and the definitions of
the other symbols are the same as those in Tab. \ref{tab:crs3773}.}
\begin{center}
\begin{tabular}{l|c|c|c|c|c|c|c|c}
\hline $e^+e^-\to f$ &$N^{\pi^0_{\gamma_i\gamma_j}}_{\rm obs}$
& $N^{\rm rc}$ & $N^{\pi^0_{\gamma_i\gamma_j}}_{\rm sid}$ & $N^{\rm b}$ &
$N^{\rm net}$ & $\epsilon(\%)$ & $\Delta_{\rm sys}(\%)$ & $\sigma^{\rm obs}$ [pb] \\
\hline
$\pi^+\pi^-\pi^0\pi^0$ & $132.6\pm15.4$ & $8.0\pm2.8$ & $6.6\pm7.1$ & $2.3\pm0.8$ & $115.7\pm17.2$ & $6.24\pm0.23$ & 13.5 & $292.2\pm43.4\pm39.4$\\
$K^+K^-\pi^0\pi^0$ & $12.0\pm3.9$ & $0.0\pm0.0$ & $0.8\pm1.4$ & $1.0\pm0.3$ & $10.2\pm4.2$ & $2.96\pm0.14$ & 14.8 & $54.3\pm22.1\pm8.0$\\
$2(\pi^+\pi^-\pi^0)$ & $147.6\pm16.2$ & $18.3\pm8.6$ & $8.7\pm7.7$ & $0.3\pm0.2$ & $120.3\pm19.9$ & $1.76\pm0.07$ & 15.3 & $1077.3\pm178.1\pm164.8$ \\
$K^+K^-\pi^+\pi^-\pi^0\pi^0$ & $25.7\pm5.8$ & $7.0\pm2.6$ & $0.7\pm1.5$ & $0.2\pm0.1$ & $17.8\pm6.5$ & $0.74\pm0.03$ & 15.1 & $379.1\pm139.1\pm57.2$ \\
$3(\pi^+\pi^-)\pi^0\pi^0$ & $25.7\pm6.4$ & $7.0\pm2.6$ & $0.0\pm0.0$ & $0.0\pm0.0$ & $18.7\pm6.9$ & $0.35\pm0.02$ & 19.1 & $842.1\pm311.1\pm160.8$ \\
\hline
\end{tabular}
\label{tab:crs3650}
\end{center}
\end{table*}

\begin{table*}[htbp]
\caption{The observed cross sections for $e^+e^-\to$ exclusive light
hadrons at $\sqrt s = 3.6648$ GeV, where
$N^{\pi^0_{\gamma_i\gamma_j}}_{\rm obs}$ is the fitted number of events
with $M_{\gamma_{i'}\gamma_{j'}}$ in the $\pi^0$ signal region
observed from the continuum data 2, $N^{\rm up}$ is the
upper limit on the number of the signal events, $\sigma^{\rm up}$ is
the upper limit on the observed cross section set at 90\% C.L., and
the definitions of the other symbols are the same as those in Tab.
\ref{tab:crs3773}.}
\begin{center}
\begin{tabular}{l|c|c|c|c|c|c|c|c}
\hline $e^+e^-\to f$ &$N^{\pi^0_{\gamma_i\gamma_j}}_{\rm obs}$
& $N^{\rm rc}$ & $N^{\pi^0_{\gamma_i\gamma_j}}_{\rm sid}$
& $N^{\rm b}$ & $N^{\rm net}$ (or $N^{\rm up}$ )& $\epsilon(\%)$ &
$\Delta_{\rm sys}(\%)$& $\sigma^{\rm obs}$
(or $\sigma^{\rm up}$) [pb] \\
\hline
$\pi^+\pi^-\pi^0\pi^0$      & $24.9\pm5.3$ & $1.0\pm1.0$ & $0.0\pm0.0$ & $0.4\pm0.1$ & $23.5\pm5.4$ &$6.06\pm0.20$ & 12.8 & $397.3\pm91.3\pm50.9$\\
$K^+K^-\pi^0\pi^0$          & 1 &    -     &    -     &    -     & $<4.36$ & $3.14\pm0.14$ & 12.0 & $<161.7$\\
$2(\pi^+\pi^-\pi^0)$        & $16.9\pm4.8$ & $2.0\pm1.4$ & $0.0\pm0.0$ & $0.1\pm0.1$ & $14.8\pm5.0$ & $1.73\pm0.06$ & 13.9 & $876.4\pm296.1\pm121.8$ \\
$K^+K^-\pi^+\pi^-\pi^0\pi^0$& 2 &    -     &    -     &    -     & $<5.91$ & $0.77\pm0.03$ & 15.1 & $<926.2$ \\
$3(\pi^+\pi^-)\pi^0\pi^0$   & 3 &    -     &    -     &    -     & $<7.42$ & $0.35\pm0.02$ & 17.2 & $<2623.1$ \\
\hline
\end{tabular}
\label{tab:crs36648}
\end{center}
\end{table*}

\begin{table*}[htbp]
\caption{The upper limit on the observed cross section
$\sigma^{\rm up}_{\psi(3770)\to f}$ at $\sqrt s=3.773$ GeV and
the upper limit on the branching fraction
${\mathcal B}^{\rm up}_{\psi(3770)\to f}$ for
$\psi(3770)\to f$ set at 90\% C.L.. The $\sigma_{\psi(3770)\to f}$
in the second column is calculated with Eq. (\ref{eq:obscrs}), where
the first error is the statistical, the second is the independent
systematic, and the third is the common systematic error.}
\begin{center}
\begin{tabular} {lccl} \hline
Decay Mode &$\sigma_{\psi(3770)\to f}$ &$\sigma^{\rm
up}_{\psi(3770)\to f}$
&$B^{\rm up}_{\psi(3770)\to f}$\\
&[pb]&[pb]&[$\times$10$ ^{-3}$] \\ \hline
$\pi^+\pi^-\pi^0\pi^0$      &$-58.6\pm47.7\pm25.7\pm6.5$   &$<61.1 $&$<8.9$\\
$K^+K^-\pi^0\pi^0$          &$-24.1\pm24.1\pm5.8\pm2.7$    &$<28.9 $&$<4.2 $\\
$2(\pi^+\pi^-\pi^0)$        &$43.3\pm204.9\pm95.4\pm5.7$   &$<399.5$&$<58.5$\\
$K^+K^-\pi^+\pi^-\pi^0\pi^0$&$-168.8\pm156.2\pm32.9\pm22.3$&$<182.1$&$<26.7$\\
$3(\pi^+\pi^-)\pi^0\pi^0$
&$185.7\pm392.2\pm122.0\pm29.9$&$<801.6$&$<117.4$\\\hline
\end{tabular}
\label{tab:up_psipp}
\end{center}
\end{table*}


\begin{thebibliography}{}

\bibitem{pdg04} Particle Data Group, S. Eidelman $et$ $al.$,
Phys. Lett. B 592 (2004) 1.
\bibitem{hepex_0506051} G. Rong, D. H. Zhang and J. C. Chen, hep-ex/0506051.
\bibitem{prl96_092002} CLEO Collaboration, D. Besson, $et$ $al.$,
Phys. Rev. Lett. 96 (2006) 092002.
\bibitem{plb641_145} BES Collaboration, M. Ablikim $et$ $al.$,
Phys. Lett. B 641 (2006) 145.
\bibitem{prl97_121801} BES Collaboration, M. Ablikim $et$ $al.$,
Phys. Rev. Lett. 97 (2006) 121801.
\bibitem{plb659_74} BES Collaboration, M. Ablikim $et$ $al.$,
\bibitem{prd76_000000} BES Collaboration, M. Ablikim $et$ $al.$,
Phys. Rev. D 76 (2007) 122002.
\bibitem{pdg07} Particle Data Group, 2007 partial updata for edition 2008
(URL:http://pdg.lbl.gov).
\bibitem{prl101_102004} BES Collaboration, M. Ablikim $et$ $al.$,
Phys. Rev. Lett. 101 (2008) 102004.
\bibitem{plb668_263} BES Collaboration, M. Ablikim $et$ $al.$,
Phys. Lett. B 668 (2008) 263.

\bibitem{hepnp28_325} BES Collaboration, J. Z. Bai $et$ $al.$, High Energy Physics and Nuclear Physics 28(4)
(2004) 325.
\bibitem{plb605_63} BES Collaboration, M. Ablikim $et$ $al.$, Phys. Lett. B 605
(2005) 63.

\bibitem{prl96_082004} CLEO Collaboration, N. E. Adam $et$ $al.$, Phys.
Rev. Lett. 96 (2006) 082004.
\bibitem{prl96_182002} CLEO Collaboration, T. E. Coans $et$ $al.$, Phys.
Rev. Lett. 96 (2006) 182002.
\bibitem{prd74_031106} CLEO Collaboration, B. A. Briere $et$ $al.$,
Phys. Rev. D 74 (2006) 031106.
\bibitem{prd74_012005} CLEO Collaboration, D. Cronin-Hennessy $et$ $al.$,
Phys. Rev. D 74 (2006) 012005.

\bibitem{prd70_077101} BES Collaboration, M. Ablikim $et$ $al.$, Phys. Rev. D 70,
(2004) 077101.
\bibitem{prd72_072007} BES Collaboration, M. Ablikim $et$ $al.$, Phys. Rev. D 72,
(2005) 072007.
\bibitem{plb650_111} BES Collaboration, M. Ablikim $et$ $al.$,
Phys. Lett. B 650 (2007) 111.
\bibitem{plb656_30} BES Collaboration, M. Ablikim $et$ $al.$,
Phys. Lett. B 656 (2007) 30.
\bibitem{epjc52_805} BES Collaboration, M. Ablikim $et$ $al.$,
Eur. Phys. J. C 52 (2007) 805.

\bibitem{prl96_032003} CLEO Collaboration, G. S. Huang $et$ $al.$, Phys.
Rev. Lett. 96 (2006) 032003.
\bibitem{prd73_012002} CLEO Collaboration, G. S. Adams $et$ $al.$, Phys.
Rev. D 73 (2006) 012002.

\bibitem{nima344_319} BES Collaboration, J. Z. Bai $et$ $al.$,
Nucl. Instrum. Methods A 344 (1994) 319.
\bibitem{nima458_627} BES Collaboration, J. Z. Bai $et$ $al.$,
Nucl. Instrum. Methods A 458 (2001) 627.
\bibitem{plb597_39} BES Collaboration, M. Ablikim $et$ $al.$,
Phys. Lett. B 597 (2004) 39.

BES Collaboration, M. Ablikim $et$ $al.$, Phys. Lett. B 603 (2004)
130.

BES Collaboration, M. Ablikim $et$ $al.$, Phys. Lett. B 608 (2005)
24.

\bibitem{npb727_395} BES Collaboration, M. Ablikim $et$ $al.$,
Nucl. Phys. B 727 (2005) 395.

\bibitem{prd57_3873} G. J. Feldman and R. D. Cousins,
Phys. Rev. D 57 (1998) 3873.

\bibitem{yf41_377} E. A. Kuraev and V. S. Fadin,
Yad. Fiz. 41 (1985) 377.
\bibitem{cpc79_291} E. Barberio and Z. Was,
Comput. Phys. Commun. 79 (1994) 291.

\bibitem{nima552_344} BES Collaboration, M. Ablikim $et$ $al.$,
Nucl. Instrum. Methods A 552 (2005) 344.

\bibitem{plb652_238} BES Collaboration, M. Ablikim $et$ $al.$,
Phys. Lett. B 652 (2007) 238.

\end{thebibliography}
\end{document}